# RESONANT FREQUENCY CONTROL FOR THE PIP-II INJECTOR TEST RFQ: CONTROL FRAMEWORK AND INITIAL RESULTS*

A. L. Edelen[†], S. G. Biedron, S. V. Milton, Colorado State University, Fort Collins, CO, USA
D. Bowring, B. E. Chase, J. P. Edelen, D. Nicklaus, J. Steimel, Fermilab, Batavia, IL, USA

## Abstract

For the PIP-II Injector Test (PI-Test) at Fermilab, a four-vane radio frequency quadrupole (RFQ) is designed to accelerate a 30-keV, 1-mA to 10-mA, $H^-$ beam to 2.1 MeV under both pulsed and continuous wave (CW) RF operation. The available headroom of the RF amplifiers limits the maximum allowable detuning to 3 kHz, and the detuning is controlled entirely via thermal regulation. Fine control over the detuning, minimal manual intervention, and fast trip recovery is desired. In addition, having active control over both the walls and vanes provides a wider tuning range. For this, we intend to use model predictive control (MPC). To facilitate these objectives, we developed a dedicated control framework that handles higher-level system decisions as well as executes control calculations. It is written in Python in a modular fashion for easy adjustments, readability, and portability. Here we describe the framework and present the first control results for the PI-Test RFQ under pulsed and CW operation.

## INTRODUCTION

The resonant frequency of the RFQ may be maintained despite changes in RF heating through thermal control. For the PI-Test RFQ, we use an internal water-cooling system [1]. Both the thermal time constants and transport delays present in such systems limit the efficacy of standard PI control. The control problem is further complicated by the cavity geometry: different rates of thermal expansion and contraction of the main internal components (the walls and the vanes/pole tips) result in a large transient frequency response under changes in average RF power. At present, the resonant frequency of other RFQs is regulated with a PI loop around the vanes, while the walls are held constant [2,3]. In contrast, a joint control loop that governs both the wall and vane temperatures enables simultaneous exploitation of their individual impacts on the resonant frequency. As discussed previously [1,4,5], these system characteristics motivate the use of MPC.

In support of this, a dedicated control framework was developed to handle high-level decisions and execute control calculations. Because multiple operational modes are required, the framework is written in Python and in a modular fashion to facilitate easy modifications to the code. The framework interfaces with ACNET (Fermilab's main control system) and the RFQ/cooling system via a custom protocol generated with a novel protocol compiler [6,7]. This framework is operational for the RFQ and could be modified for similar control tasks at Fermilab.

In this paper we describe the operational needs for the RFQ, the design of the control framework, and initial control results. This work represents a first test of resonant frequency control over the RFQ, a first test of the framework, and a first test of using a dedicated Python program at Fermilab interfaced with the main control system via the protocol compiler.

## SYSTEM DESCRIPTION AND OPERATIONAL GOALS

More details on the system and control challenges therein are described in [1, 5]. The low-level radio frequency (LLRF) system can compensate for detuning of the cavity only up to 3 kHz by taking advantage of the available overhead in the RF power amplifiers. This limitation translates directly to challenges for the resonance control system: detuning beyond 3 kHz occurs rapidly under changes in average RF power, particularly in CW mode, due to the frequency response of the vanes [8, 9]. In addition, the RFQ operates in both CW and pulsed RF modes, resulting in variable RF heating. Other challenges are imposed by the architecture of the water-cooling system. Transport delays and thermal time constants result in open-loop settling times on the order of tens of minutes during normal operation (e.g. see Figure 1). Finally, the coupling between the wall and vane circuits, the transient frequency response, the nonlinear valve flow curves, and fluctuations in the temperature of the cold water supply make the system more difficult to control.

The LLRF system [10] is capable of operating in either SEL mode (in which the drive frequency follows the cavity resonant frequency) or GDR mode (in which the drive frequency is set). In SEL mode, the use of RF overhead is minimized due to the changing of the drive frequency to match the RFQ, thereby also minimizing the reflected power. As such it is useful to switch into SEL mode automatically when the detuning increases beyond a tolerable threshold.

Additionally, accommodation of multiple control algorithms is desired. MPC frequency control will be the main method; however, PI frequency control using the vane valve is also desired as an auxiliary mode. In addition, for fast trip recovery it is useful to control the water temperature directly. Next, another desired mode is control of the RF forward power magnitude during a cold start or recovery from a trip. This would start out as a simple ramp, but could eventually be incorporated into an MPC routine.

In the event of an RF trip, the required recovery time for the RFQ is no more than 10x the length of the trip, with a target requirement of 2x the duration of the trip. These constraints and desired system flexibility motivate the development of a modular control system architecture that can be easily modified and would facilitate the use of multiple control algorithms, including MPC.

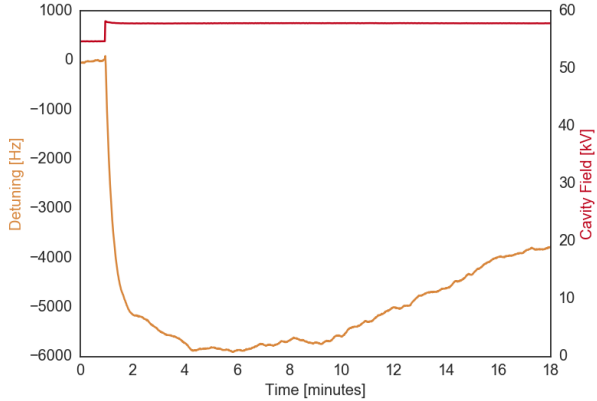

Figure 1: Example of uncontrolled detuning in CW mode under a small change in cavity field (55 kV to 58 kV).

## FRAMEWORK DESCRIPTION

### Framework Tasks

The control framework handles a variety of high-level decisions related to RFQ operation, including: switching between SEL and GDR mode automatically based on reflected power, measured detuning, and cavity field setpoint vs. read-back; setting the LLRF averaging windows; switching between user-requested operational states; detecting RF trips and taking appropriate action (e.g. water temperature control) until power returns; and calculating the resonant frequency using a method appropriate for the current operational state (i.e. SEL vs. GDR mode).

### Framework Interfaces

The controller communicates with several other subsystems, including LLRF, a Cryo-con 18$i$ temperature monitor [11], and a programmable logic computer (PLC) for the water system. All of these communications occur through the ACNET control system via an ACNET Erlang front-end. This front-end is connected to the resonance controller via a User Datagram Protocol (UDP). We use a custom protocol (i.e. specific to this controller) generated with Fermilab's protocol compiler [6,7]. The front-end then handles the communication with the variety of other subsystems through standard ACNET messages, thus hiding that complexity from the controller

An operator can specify the desired operational mode as an input to the controller. Error read-backs, a heartbeat, and the present operational mode are sent back from the framework to the user. From the LLRF system, readings include the present SEL/GDR state, an indicator for pulsed or CW mode, the forward and cavity phase readings, RF repetition rate and pulse length, forward and reflected RF power magnitudes, cavity field magnitude, and RF timing and averaging windows. The Cryo-con 18$i$ and PLC return temperature sensor readings, pressure and flow readings, and flow control valve readings. The controller sets the LLRF system to SEL or GDR mode and sets appropriate window settings. For the PLC, the resonance controller sets the flow control valve apertures. Figure 2 shows a simplified view of the system interfaces.

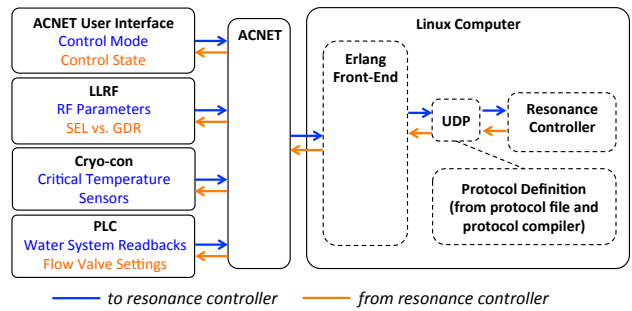

Figure 2: Simplified view of the system interfaces.

### Framework Structure

The framework itself consists of three main units: a main program, an action module, and a control module (see Figure 3). The control module contains the control algorithms and their helper functions. The action module contains several action functions with computations and checks that need to be completed for different operational modes. The main program includes execution of initialization tasks, interpretation of user requests, receiving of inputs from the front-end, selection of action module functions based on user requests and readings from the machine, and collection and sending of commands to the front-end.

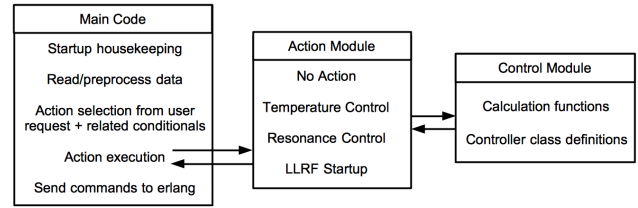

Figure 3: Modular control framework structure.

### Main Program Flow

The main program flow is as follows. (1) Upon startup, the program builds a data buffer for MPC, sets the LLRF windows, and initializes variables. (2) At each execution cycle of the program, it gathers new readings, checks the requested state, pre-processes some of the data, makes relevant calculations (e.g. resonant frequency), and selects the appropriate action function. (3) The action is executed, which also involves calling the control module when applicable. Mode-specific settings are then sent back to the main program. (4) Finally, the information is collected and commands are sent to the Erlang front-end. At present, steps 2-4 repeat at a 1 Hz rate, but this rate may be increased in the future if need be. Figure 4 shows the program flow for resonance control specifically.

## INITIAL CONTROL RESULTS

Recently, the main framework was commissioned and PI control over the frequency in both pulsed and CW mode was tested. This included frequency recovery after RF trips, user-driven state switching, and automated switching between SEL and GDR modes. At present, RF recovery after a trip (that is, returning RF power to its original level) is conducted manually.

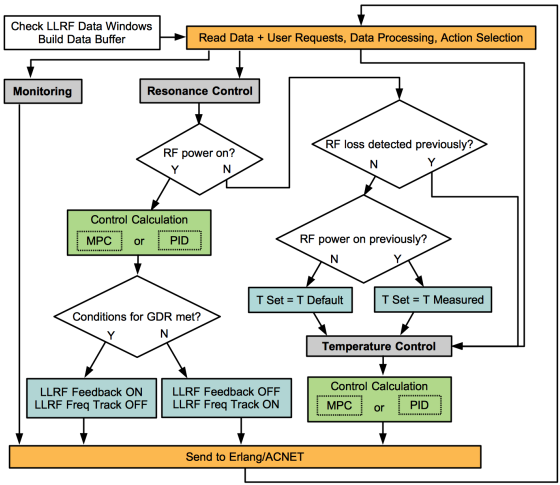

Figure 4: Overview of the program flow.

The PI controller regulates the resonant frequency of the RFQ by adjusting the vane flow valve aperture. Because the flow responses of the valves are coupled and nonlinear [1], a weighted correction is applied to the PI control output. The correction is inversely proportional to predicted change in flow for a given change in valve setting. Because of the transport delays in the system, a corresponding delay is placed into the frequency and temperature PI controllers. The weighting of the PI gains, the flow valve correction, and amount of delay were all tuned during commissioning.

Figures 5, 6, and 7 show examples of the PI frequency controller performance. In Figure 5, the pulse duration was increased by 2 ms while the cavity field was at 65 kV. Figure 6 shows frequency control under CW operation for a similar step in cavity field to that shown in Figure 1. Figure 7 shows frequency recovery from a 10-second trip at full specified field (60 kV), in which the detuning is reduced to 3 kHz in 140 seconds. As such, PI control does not fully meet the specification for trip recovery time at full field, although frequency trips at lower field values can be recovered in under 10x the trip duration. Operational experience indicates that PI control may be brought into specification by slightly delaying the activation of the PI loop after power is restored.

## CONCLUSION AND FURTHER WORK

PI resonant frequency control over the RFQ makes stable CW and pulsed operation in GDR mode possible. The next step of this work is to commission MPC. This should provide compensation for water temperature fluctuations, will enable finer control over the resonant frequency by leveraging both the wall and vane responses, and should improve the frequency recovery time after a trip. Automated RF recovery will also be added.

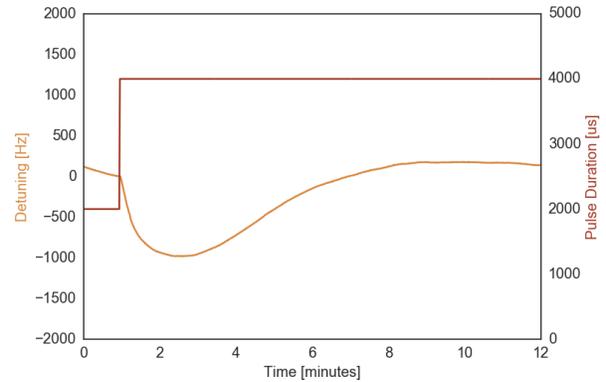

Figure 5: PI frequency control during pulsed RF operation for a 2-ms increase in pulse duration and a cavity field of 65 kV.

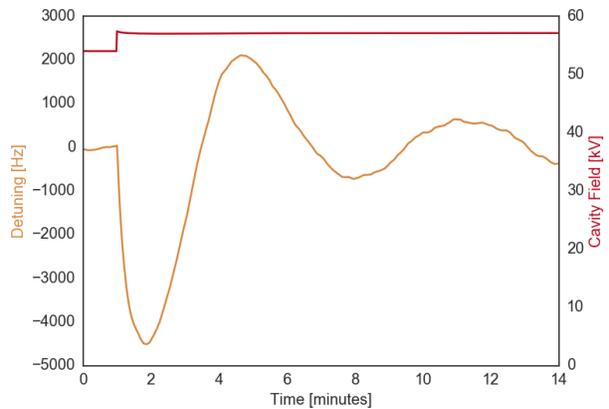

Figure 6: Frequency control under CW operation under a small change in cavity field. This is roughly the same step size as used for the uncontrolled response shown in Figure 1.

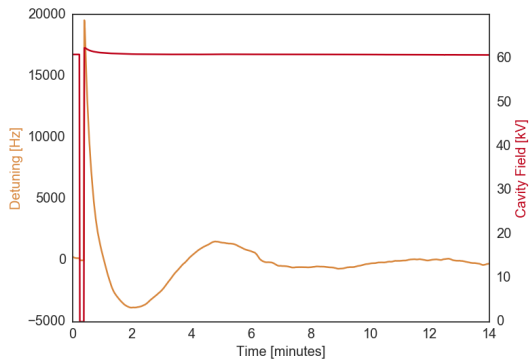

Figure 7: Frequency recovery from a 10-second RF trip at 60 kV using PI control over the vane flow valve.

## ACKNOWLEDGMENTS

The authors thank R. Neswold and C. King for their work on the protocol compiler, as well as C. Baffes, M. Ball, J.-P. Carneiro, E. Cullerton, J. Czaikowski, B. Hanna, R. Pasquinelli, D. Peterson, G. Romanov, A. Shemyakin, P. Varghese, and T. Zuchnik.

# REFERENCES

[1] D. Bowring *et al.,* "Resonance Control for Fermilab's PXIE RFQ," in *Proc. IPAC'16*, Busan, Korea, May 2016, paper MOPMW026.

[2] P. N. Ostruomov *et al.*, "Development and beam test of a continuous wave radio frequency quadrupole accelerator," *PRST-AB* 15, 110101, 2012.

[3] W. Jing *et al.,* "Multi-physics analysis of the RFQ for injections scheme II of C-ADS driver linac," *Chinese Physics C* 38, 10, 2014.

[4] A. L. Edelen *et al.,* "Neural Networks for Modeling and Control of Particle Accelerators," *IEEE Transactions on Nuclear Science,* vol. 63, no. 2, Apr. 2016.

[5] A. L. Edelen *et al.,* "Neural Network Model of the PXIE RFQ Cooling System and Resonant Frequency Response," in *Proc. IPAC'16,* Busan, Korea, May 2016, paper THPOY020.

[6] R. Neswold, C. King, "Generation of Simple, Type-Safe Messages for Inter-task Communications," in *Proc. ICALEPCS 2009*, Kobe, Japan, Oct., 2009.

[7] R. Neswold, C. King, "Further Developments in Generating Type-Safe Messaging," in *Proc. ICALEPCS 2011*, Grenoble, France, Oct. 2011.

[8] A. R. Lambert, "FNAL PXIE 162.5 MHz RFQ GENERAL RF," LBNL Eng. Note 10773, 11 Jan 2013.

[9] A. R. Lambert *et al.*, "High-Intensity Proton RFQ Accelerator Fabrication Status for PXIE," in *Proc. IPAC'15,* Richmond, VA, May 2015, paper WEPTY045.

[10] J. P. Edelen *et al*, "Low level RF control for the PIP-II Injector Test RFQ", *NAPAC 2016*, Chicago, IL, this conference.

[11] *User's Guide, Model 18i/14i/M12i Cryogenic Temperature Monitors, Revision 3c*. Cryogenic Control Systems, Inc., 2016.